\documentclass[showpacs,aps,prb,twocolumn]{revtex4}
\usepackage{amsmath}
\usepackage{amsfonts}
\usepackage{amssymb}
\usepackage{amsthm}
\usepackage{graphicx}
\usepackage{color}
\begin{document}


\title{Classical and Quantum Polyhedra}
\author{John Schliemann}
\affiliation{ Institute for Theoretical Physics, University of
Regensburg, D-93040
Regensburg, Germany}
\date{\today}

\begin{abstract}
Quantum polyhedra constructed from angular momentum operators are
the building blocks of space in its quantum description as advocated by
Loop Quantum Gravity.
Here we extend previous results on the semiclassical properties of quantum
polyhedra. Regarding tetrahedra, we compare the results from
a canonical quantization of the classical system with a recent
wave function based
approach to the large-volume sector of the quantum system.
Both methods agree in the leading order of the resulting
effective operator (given by an harmonic oscillator), while minor
differences occur in higher corrections. Perturbative inclusion of such
corrections improves the approximation to the eigenstates.
Moreover, the comparison of both methods
leads also to a full wave function description of the eigenstates of the
(square of the) volume operator at negative eigenvalues of large modulus.

For the case of general quantum polyhedra described by discrete angular momentum
quantum numbers we formulate a set of
quantum operators fulfilling in the semiclassical regime 
the standard commutation relations
between momentum and position. Differently from previous formulations,
the position variable here is chosen to have dimension of (Planck) length
squared which facilitates the identification of quantum corrections. 
Finally, we provide expressions for the pentahedral volume in terms of
Kapovich-Millson variables.
\end{abstract}
\maketitle

\section{introduction}

The quantum volume operator is among the most intensively investigated
items in the field
of Loop Quantum Gravity and pivotal for the construction
of space-time dynamics within this theoretical framework 
\cite{Rovelli04,Thiemann07,Perez13}. 
Traditionally two versions
of such an operator are discussed, due to Rovelli and Smolin \cite{Rovelli95},
and to Ashtekar and Lewandowski \cite{Ashtekar95}, respectively,
and considerable attention has been devoted to their properties and 
interrelations\cite{Loll95,Loll96,DePietri96,DePietri97,Thiemann98,Barbieri98,Carbone02,Brunnemann06,Giesel06a,Giesel06b,Meissner06,Brunnemann08a,Brunnemann08b,Dittrich09}.
More recently, Bianchi, Dona, and Speziale \cite{Bianchi11a} offered a 
third proposal for a volume operator which is closer to the concept of spin 
foams \cite{Perez13}. It relies on
an older geometric theorem due to Minkowski \cite{Minkowski1897} stating
that $N$ face areas $A_i$, $i\in\{1,\dots,N\}$, with normal vectors
$\vec n_i$ such that 
\begin{equation}
\sum_{i=1}^N\vec A_i=0
\label{clorel}
\end{equation}
for $\vec A_i=\vec n_iA_i$ uniquely define a convex polyhedron of $N$ faces
with areas $A_i$.
The approach of Ref.\cite{Bianchi11a} amounts in expressing the
volume of a classical polyhedron in terms of its face areas, which are in turn 
promoted to be operators.
Minkowski\rq s proof, however, is not constructive, and a remaining
obstacle of the above route to a volume operator is to actually find the
shape of a general polyhedron given its face areas and face normals
\cite{Bianchi11a,Bianchi11b,Bianchi12,Haggard13,Coleman-Smith13}.

Such difficulties do not occur in the simplest case of a polyhedron, i.e.
a tetrahedron consisting
of four faces represented by angular momentum operators coupling
to a total spin singlet \cite{Barbieri98,Baez99}. 
Indeed, for such a quantum tetrahedron all three definitions of the volume
operator coincide. On the other hand, for a classical
tetrahedron the general phase space
parametrization devised by Kapovich and Millson \cite{Kapovich96}
results in just one pair of canonical variables, and the (square of the)
volume operator can explicitly formulated in terms of these quantities
\cite{Bianchi12,Aquilanti13}.
Moreover , Bianchi and
Haggard have performed a Bohr-Sommerfeld quantization of the classical
tetrahedron where the role of an Hamiltonian generating classical orbits
is played by the volume operator squared.
The resulting semiclassical eigenvalues agree
extremely well with exact numerical data \cite{Bianchi11b,Bianchi12}.
The above observations make clear that classical tetrahedra, arguably the
simplest structures a volume can be ascribed to, should be considered
as perfectly integrable systems. In turn, a quantum tetrahedron can be
viewed as the ``hydrogen atom'' of quantum spacetime, whereas the next
complicated case of a pentahedron might be referred to as the
``helium atom'' \cite{Coleman-Smith13}.

Most recently, the present author has put forward yet another approach to the
semiclassical regime of quantum tetrahedra \cite{Schliemann13}. 
Here, by combining
observations on the volume operator squared and its eigenfunctions
(as opposed to the eigenvalues), an effective operator in terms of a
quantum harmonic oscillator was derived providing an accurate
as well as transparent description of the the large-volume
sector. One of the purposes of the present work is to demonstrate the
relation between the different treatments of quantum tetrahedra sketched
above.
 
The outline of this paper is as follows. In section \ref{clpoly} we
first summarize the Kapovich-Millson phase space parametrization of general
classical polyhedra (section \ref{clpolyKP}) before reviewing and extending
in section \ref{cltetra}
results for the classical tetrahedron. In particular, we derive an expansion
of the volume squared around its maximum and minimum in up to 
quadrilinear order. Section \ref{quanttetra} is devoted to the quantum
tetrahedron. We first outline in section \ref{quanttetragen}
elementary facts about the volume operator
and its Hilbert space, and we point out several relations between appropriate
quantum operators  which have analogs in the classical tetrahedron.
Next the analysis of Ref.~\cite{Schliemann13} is extended to higher
corrections to the resulting harmonic oscillator of up to fourth order. The
results are compared with the outcome of a canonical quantization
of the classical volume expression. The expression of the pentahedral volume
in terms of Kapovich-Millson variables is discussed in appendix \ref{penta}.
Finally we construct in section \ref{quantpolygen} 
a set of quantum operators for general polyhedra whose commutation relations
approach in the semiclassical limit
the standard commutators between momentum and position.

\section{Classical Polyhedra}
\label{clpoly}

Let us first recall the essentials of the polyhedral phase space 
parametrization due to Kapovich and Millson \cite{Kapovich96}.

\subsection{Kapovich-Millson Phase Space Variables}
\label{clpolyKP}

Viewing the vectors $\vec A_i$ as angular momenta, the Poisson bracket
of arbitrary functions of these variables read
\begin{equation}
\left\{f,g\right\}=\sum_{i=0}^N\vec A_i\cdot
\left(\frac{\partial f}{\partial\vec A_i}
\times\frac{\partial g}{\partial\vec A_i}\right)\,.
\end{equation}
In order to implement the closure relation (\ref{clorel}) one defines
\begin{equation}
\vec p_i=\sum_{j=1}^{i+1}\vec A_j
\end{equation}
for $i\in\{1,\dots,N-3\}$ resulting in $N-3$ momenta
$p_i=|\vec p_i|$. Defining now 
\begin{equation}
\vec v_i=\vec p_i\times\vec A_{i+1}\qquad,\qquad
\vec w_i=\vec p_i\times\vec A_{i+2}\,,
\label{defvw}
\end{equation}
such that $\vec v_{i+1}=\vec w_i$ ($i<N-3$) and
\begin{equation}
\vec p_i\cdot\vec v_i=\vec p_i\cdot\vec w_i=0\,,
\end{equation}
the canonical conjugate variables
$q_i$ are then given be the angle between $\vec v_i$, $\vec w_i$. Indeed, 
and a straightforward calculation shows that these quantities fulfill
indeed the canonical Poisson relations \cite{Kapovich96,Bianchi12}.
\begin{equation}
\left\{p_i,q_j\right\}=\delta_{ij}\,.
\end{equation}

\subsection{The Tetrahedron}
\label{cltetra}

The classical volume of a tetrahedron can be expressed as
\begin{equation}
V=\frac{\sqrt{2}}{3}\sqrt{|\vec A_1\cdot(\vec A_2\times\vec A_3)|}
\label{classV}
\end{equation}
suggesting to investigate the quantity
\begin{equation}
Q=\vec A_1\cdot(\vec A_2\times\vec A_3)\,.
\label{classQ}
\end{equation}
The latter can indeed easily be expressed in terms of the
phase space variables $p_1$, $q_1$ using the observation \cite{Bianchi12}
\begin{equation}
\vec v_1\times\vec w_1=Q\vec p_1\,.
\label{clvxw}
\end{equation}
Moreover, it is easily seen that
\begin{equation}
|\vec v_1|=|\vec A_1\times\vec A_2|=2\Delta(A_1,A_2,p_1)
\label{clvsq}
\end{equation}
where $\Delta(a,b,c)$ is the area of a triangle with edges $a,b,c$ expressed
via Heron\rq s formula,
\begin{equation}
\Delta(a,b,c)=\frac{1}{4}\sqrt{\left((a+b)^2-c^2\right)
\left(c^2-(a-b)^2\right)}\,.
\label{defDel}
\end{equation}
Analogously, using the closure relation (\ref{clorel}),
\begin{equation}
|\vec w_1|=|\vec A_3\times\vec A_4|=2\Delta(A_3,A_4,p_1)
\label{clwsq}
\end{equation}
such that
\begin{equation}
Q=4\frac{\Delta(A_1,A_2,p_1)\Delta(A_3,A_4,p_1)}{p_1}\sin q_1\,.
\end{equation}
In order to make closer contact to the quantum tetrahedron to be discussed
below, let us introduce the notation
\begin{equation}
A:=p_1\quad,\quad p:=-q_1+\frac{\pi}{2}
\end{equation}
fulfilling $\{p,A\}=1$ and
\begin{equation}
Q=2\tilde\beta(A)\cos p
\label{clQ}
\end{equation}
with
\begin{equation}
\tilde\beta(A)=2\frac{\Delta(A_1,A_2,A)\Delta(A_3,A_4,A)}{A}\,,
\label{clbeta}
\end{equation}
where $A$ varies according to $A^{\rm min}\leq A\leq A^{\rm max}$ with
\begin{eqnarray}
A^{\rm min} & = & \max\{|A_1-A_2|,|A_3-A_4|\}\,,
\label{cltetkcond1}\\
A^{\rm max} & = & \min\{A_1+A_2,A_3+A_4\}\,.
\label{cltetkcond2}
\end{eqnarray}
An expression close to (\ref{clQ}) was also found in Ref.~\cite{Aquilanti13}
in the semiclassical limit of a quantum tetrahedron.
 
Obviously, $\tilde\beta(A)$ is a nonnegative 
function with $\tilde\beta(A^{\rm min})=\tilde\beta(A^{\rm max})=0$, and 
it is not difficult to verify that it has a unique maximum at 
some $A=\bar A$ between $A^{\rm min}$ and
$A^{\rm max}$ \cite{Schliemann13}. 
Thus, $Q$ has a unique maximum at $A=\bar k$ and $p=0$ while
the unique minimum lies at $p=\pi$.
Expanding around the maximum gives ($x:=A-\bar A$, $|x|\ll 1$, $|p|\ll 1$)
\begin{eqnarray}
Q(p,x) & = & \tilde q\Biggl[1-\frac{p^2}{2}-\frac{\tilde\omega^2}{2}x^2
+\frac{\tilde c}{3}x^3+\frac{\tilde d}{4}x^4\nonumber\\
& & \quad+\frac{\tilde\omega^2}{4}x^2p^2+\frac{p^4}{24}+\cdots\Biggr]
\label{qexpmax}
\end{eqnarray} 
with
\begin{equation}
\tilde q=2\tilde\beta(\bar A)\quad,\quad
\tilde \omega^2=-\frac{\left(\frac{d^2\tilde\beta(A)}{dA^2}\right)_{A=\bar A}}
{\tilde\beta(\bar A)}>0
\end{equation}
and 
\begin{equation}
\tilde c=\frac{\left(\frac{d^3\tilde\beta(A)}{dA^3}\right)_{A=\bar A}}
{2\tilde\beta(\bar A)}\quad,\quad
\tilde d=\frac{\left(\frac{d^4\tilde\beta(A)}{dA^4}\right)_{A=\bar A}}
{6\tilde\beta(\bar A)}\,.
\end{equation}
The analogous expansion around the minimum reads 
($p=\pi+(p-\pi)$, $|p-\pi|\ll 1$)
\begin{eqnarray}
Q^{\prime}(p,x) & = & -\tilde q\Biggl[1-\frac{(p-\pi)^2}{2}
-\frac{\tilde\omega^2}{2}x^2
+\frac{\tilde a}{3}x^3+\frac{\tilde b}{4}x^4\nonumber\\
& & \quad+\frac{\tilde\omega^2}{4}x^2(p-\pi)^2
+\frac{(p-\pi)^4}{24}+\cdots\Biggr]\,.
\label{qexpmin}
\end{eqnarray} 
Concentrating in both cases on the quadratic contributions, one obtains
two harmonic oscillators,
\begin{eqnarray}
Q_{\rm osc}(p,x) & = & 
\tilde q\left[1-\frac{p^2}{2}-\frac{\tilde\omega^2}{2}x^2\right]\,,
\label{oscmax}\\
Q_{\rm osc}^{\prime}(p,x) & = & 
-\tilde q\left[1-\frac{(p-\pi)^2}{2}-\frac{\tilde\omega^2}{2}x^2\right]\,.
\label{oscmin}
\end{eqnarray} 

Finally, it is certainly desirable to also express the volume of higher
polyhedra in terms of Kapovich-Millson variables. Appendix \ref{penta}
details the case of the pentahedron. As shown there, the above task is certainly
feasible, but leads to unpleasantly complicated expressions which inhibit
analytical progress.

\section{Quantum Polyhedra}
\label{quantpoly}

\subsection{The Quantum Tetrahedron}
\label{quanttetra}

We begin by reviewing and extending general results of quantum
tetrahedra.

\subsubsection{General Properties}
\label{quanttetragen}

A quantum tetrahedron is defined by four angular momentum operators 
$\hat{\vec j}_i$,
$i\in\{1,2,3,4\}$, representing its faces and coupling to a total 
singlet\cite{Barbieri98,Carbone02,Baez99,Bianchi11b,Bianchi12}, 
i.e. the Hilbert space consists of all states $|k\rangle$
fulfilling
\begin{equation}
\left(\hat{\vec j}_1+\hat{\vec j}_2
+\hat{\vec j}_3+\hat{\vec j}_4\right)|k\rangle=0\,.
\end{equation}
A usual way to construct this space is to couple first
the pairs $\hat{\vec j}_1,\hat{\vec j}_2$ and 
$\hat{\vec j}_3,\hat{\vec j}_4$
to two irreducible SU(2) representations of dimension $2k+1$ each.
For $\hat{\vec j}_1,\hat{\vec j}_2$  this standard construction reads explicitly
\begin{equation}
\hat{\vec k}:=\hat{\vec j}_1+\hat{\vec j}_2\,,
\end{equation}
\begin{equation}
|km\rangle_{12}=\sum_{m_1+m_2=m}\langle j_1m_1j_2m_2|km\rangle
|j_1m_1\rangle|j_2m_2\rangle\,,
\end{equation}
such that
\begin{eqnarray}
\hat k^z|km\rangle_{12} & = & m|km\rangle_{12}\,,\\
\hat{\vec k}^2|km\rangle_{12} & = & k(k+1)|km\rangle_{12}\,,
\end{eqnarray}
where $\langle j_1m_1j_2m_2|km\rangle$ are Clebsch-Gordan coefficients
following their usual phase convention\cite{Edmonds57}.
Defining analogous states $|km\rangle_{34}$ for 
$\hat{\vec j}_3,\hat{\vec j}_4$, the
quantum number
$k$ becomes restricted by $k^{\rm min}\leq k\leq k^{\rm max}$ with 
\begin{eqnarray}
k^{\rm min} & = & \max\{|j_1-j_2|,|j_3-j_4|\}\,,
\label{tetkcond1}\\
k^{\rm max} & = & \min\{j_1+j_2,j_3+j_4\}\,.
\label{tetkcond2}
\end{eqnarray}
The two multiplets $|km\rangle_{12}$, $|km\rangle_{34}$ are then coupled
to a total singlet,
\begin{eqnarray}
|k\rangle & = & e^{i\frac{\pi}{2}(k-k^{\rm min})}\nonumber\\
& & \quad\cdot\sum_{m=-k}^k\frac{(-1)^{k-m}}{\sqrt{2k+1}}
|km\rangle_{12}|k(-m)\rangle_{34}\,,
\label{defsing}
\end{eqnarray}
where the phase factor in front will become useful shortly below. The states
$|k\rangle$ span a Hilbert space of dimension
$d=k^{\rm max}-k^{\rm min}+1$.

The volume operator of a quantum tetrahedron can be formulated as
\begin{equation}
\hat V=\frac{\sqrt{2}}{3}\sqrt{|\hat{\vec E}_1
\cdot(\hat{\vec E}_2\times\hat{\vec E}_3)|}
\label{quantV}
\end{equation}
where the operators
\begin{equation}
\hat{\vec E}_i=\ell_P^2\hat{\vec j}_i\,,
\label{defe}
\end{equation}
$i\in\{1,2,3,4\}$ represent the faces of the tetrahedron with
$\ell_P^2=\hbar G/c^3$ being the Planck length squared. Usually the  
operators $\hat{\vec E}_i$ are defined with additional prefactors
proportional to the Immirzi parameter on the r.h.s. of Eq.~(\ref{defe}). 
This establishes contact to the general formalism of loop quantum gravity
\cite{Rovelli04,Thiemann07,Perez13} but is unnecessary for our purposes here.
What will become important, however, is that $\ell_P^2$ is proportional
to $\hbar$.

As a result, one is led to consider the operator
\begin{equation}
\hat R=\hat{\vec j}_1\cdot(\hat{\vec j}_2\times\hat{\vec j}_3)\,,
\end{equation}
which reads in the basis of the states $|k\rangle$ as
\cite{Carbone02,Bianchi12,Levy-Leblond65,Chakrabarti64,Edmonds57,Schliemann13}
\begin{equation}
\hat R=\sum_{k=k^{\rm min}+1}^{k^{\rm max}}\alpha(k)
\left(|k\rangle\langle k-1|+|k-1\rangle\langle k|\right)
\end{equation}
with
\begin{eqnarray}
\alpha(k) & = & \frac{2}{\sqrt{k^2-1/4}}
\Delta(j_1+1/2,j_2+1/2,k)\nonumber\\
& & \quad\cdot\Delta(j_3+1/2,j_4+1/2,k)\,.
\label{alpha}
\end{eqnarray}
Note the close similarity of the expressions (\ref{alpha}) and
(\ref{clbeta}). Moreover, in the above basis $\hat Q$ couples only states
with neighboring labels and is represented by a real matrix. The latter fact
depends on the phase factor in the first line of Eq.~(\ref{defsing}). Indeed,
upon striping this factor (which is a unitary operation) $\hat R$ becomes
antisymmetric and purely imaginary. Thus, for even $d$, the eigenvalues
of $Q$ come in pairs $q,(-q)$, and since
\begin{equation}
u\hat Ru^+=-\hat R
\end{equation} 
with $u={\rm diag}(1,-1,1,-1,\dots)$, the corresponding eigenstates
$|\phi_q\rangle$, $|\phi_{-q}\rangle$ fulfill
\begin{equation}
|\phi_{-q}\rangle=u|\phi_q\rangle\,,
\label{signchange}
\end{equation} 
i.e. eigenvectors of eigenvalues differing just in sign are related to each 
other by changing the sign of any other component.
For odd $d$ an additional zero eigenvalue occurs
\cite{Brunnemann06}.

To make further contact between the classical and the quantum tetrahedron
we define in analogy to Eqs.(\ref{defvw})
\begin{eqnarray}
\hat{\vec v} & = & \frac{1}{2}\left(\hat{\vec k}\times\hat{\vec j}_2
-\hat{\vec j}_2\times\hat{\vec k}\right)=\hat{\vec j}_1\times\hat{\vec j}_2\,,\\
\hat{\vec w} & = & \hat{\vec k}\times\hat{\vec j}_3
\label{defvwop}
\end{eqnarray}
fulfilling
\begin{equation}
\frac{1}{2}\left(\hat{\vec v}\times\hat{\vec w}
-\hat{\vec w}\times\hat{\vec v}\right)
=\frac{1}{2}\left(\hat R\hat{\vec k}+\hat{\vec k}\hat R\right)\,,
\end{equation} 
which is the operator analog of Eq.~(\ref{clvxw}). Moreover, one
straightforwardly obtains
\begin{equation}
\hat{\vec v}^2=4\left(
\Delta\left(\sqrt{j_1(j_1+1)},\sqrt{j_2(j_2+1)},\hat{\vec k}\right)\right)^2
\label{quvsq}
\end{equation}
and
\begin{eqnarray}
& & \hat\Pi\hat{\vec w}^2\hat\Pi\nonumber\\
& & =4\hat\Pi\left(
\Delta\left(\sqrt{j_3(j_3+1)},\sqrt{j_4(j_4+1)},
\hat{\vec k}\right)\right)^2\hat\Pi\,,
\label{quwsq}
\end{eqnarray}
where
\begin{equation}
\hat\Pi=\sum_{k=k^{\rm min}}^{k^{\rm max}}|k\rangle\langle k|
\label{projec}
\end{equation}
is the projector onto the singlet space. Eqs.~(\ref{quvsq}),(\ref{quwsq})
are the operators analogs of Eqs.~(\ref{clvsq}),(\ref{clwsq}).

\subsubsection{Rescaling to Dimensionful Variables}
\label{quanttetrarescale}

So far we have followed the formalism common to the literature and 
parametrized the Hilbert space of the quantum tetrahedron by a dimensionless
quantum number $k$, whereas the phase space variable $A$ of the classical
tetrahedron has dimension of area. In order to establish closer contact
between both descriptions let us rescale the involved quantum numbers
by the Planck length squared according to
\begin{equation}
k \mapsto a=\ell_P^2k\quad,\quad j_i\mapsto E_i=\ell_P^2j_i
\end{equation}
to quantities having also dimension of area. As we shall see below, this
step will also provide a close analogy to standard quantum mechanics in the
Schr\"odinger representation.
The analog of the classical expression (\ref{classQ}) reads
\begin{eqnarray}
\hat Q & = & \ell_P^6\hat R
=\hat{\vec E}_1\cdot(\hat{\vec E}_2\times\hat{\vec E}_3)\\
& = & \sum_{a=a^{\rm min}+\ell_P^2}^{a^{\rm max}}\beta(a)
\left(|a\rangle\langle a-\ell_P^2|+|a-\ell_P^2\rangle\langle a|\right)
\end{eqnarray}
with
\begin{eqnarray}
\beta(a) & = & \ell_P^6\alpha(k)\\ 
& = & \frac{2}{\sqrt{a^2-\ell_p^4/4}}
\Delta(E_1+\ell_P^2/2,E_2+\ell_P^2/2,a)\nonumber\\
& & \quad\cdot\Delta(E_3+\ell_P^2/2,E_4+\ell_P^2/2,a)\,.
\label{beta}
\end{eqnarray}
The latter quantity shares the essential properties of $\tilde\beta(A)$ in
Eq.~(\ref{clbeta}). In particular $\beta(a)$ has a unique maximum at some
$a=\bar a$.

\subsubsection{Large Volumes}
\label{quanttetralarge}

In Ref.~\cite{Schliemann13} the present author has shown
how to accurately describe the large-volume (semiclassical) regime of 
$\hat Q$ (or $\hat R$)
by a quantum harmonic oscillator in real-space representation with
respect to $a$ (or $k$, respectively). Here we shall extend this analysis
taking into account higher-order corrections within the rescaled variables
introduced in the previous section.

Let us label the eigenstates of $\hat Q$ by $|n\rangle$, $n\in\{0,1,2,\dots\}$, 
in descending order of eigenvalues with $|0\rangle$ being the state
of largest eigenvalue. With respect to the basis states $|k\rangle$ they can
be expressed as 
\begin{equation} 
|n\rangle=\sum_{a=a_{\rm min}}^{a_{\rm max}}\langle a|n\rangle|a\rangle\,.
\end{equation}
Thus, taking the view of the standard Schr\"odinger formalism of elementary
quantum mechanics, the coefficients $\langle a|n\rangle$ are the  
``wave function'' of the state $|n\rangle$ with respect to the
``coordinate'' $a$. The approach of Ref.~\cite{Schliemann13} 
starts from evaluating matrix elements
\begin{eqnarray}
\langle\Phi|Q|\Psi\rangle & = & \sum_a\beta(a)
\Bigl(\langle\Phi|a\rangle\langle a-\ell_P^2|\Psi\rangle\nonumber\\
 & & \qquad+\langle\Phi|a-\ell_P^2\rangle\langle a|\Psi\rangle\Bigr)
\label{Qmat1}
\end{eqnarray}
between states lying predominantly in the 
sector of large eigenvalues by approximating the sum by an integral
introducing the integration variable $x:=a-\bar a$,
\begin{eqnarray}
\langle\Phi|Q|\Psi\rangle & \approx & 
\frac{1}{\ell_P^2}\int dx \beta(\bar a+x)
\Bigl(\tilde\Phi^*(x)\tilde\Psi(x-\ell_P^2)|\nonumber\\
 & & \qquad+\tilde\Phi^*(x-\ell_P^2)\tilde\Psi(x)\Bigr)
\label{Qmat2}
\end{eqnarray}
with $\tilde\Phi(x)=\langle\bar a+x|\Phi\rangle$,
$\tilde\Psi(x)=\langle\bar a+x|\Psi\rangle$.
Expanding now $\beta(\bar a+x)$ around its maximum at 
$\bar a$ and the wave functions
$\tilde\Phi^*(x-\ell_P^2)$, $\tilde\Psi(x-\ell_P^2)$ around $x$, one obtains
in up to fourth order in the expansions
\begin{eqnarray}
& & \langle\Phi|Q|\Psi\rangle\approx\int dx\,\Phi^*(x)
\bar q\Biggl[1-\left(-\frac{\ell_P^4}{2}\frac{d^2}{dx^2}
+\frac{\omega^2}{2}x^2\right)\nonumber\\
& & \qquad+\frac{c}{3}x^3+\frac{d}{4}x^4
-\frac{\omega^2}{8}\ell_P^4
\left(x^2\frac{d^2}{dx^2}+\frac{d^2}{dx^2}x^2\right)\nonumber\\
& & \qquad
+\frac{\ell_P^8}{24}\frac{d^4}{dx^4}\nonumber\\
& & \qquad
+\frac{\omega^2}{2}\ell_P^2\left[\frac{d}{dx},x^2\right]
+\frac{c}{3}\ell_P^2\left[\frac{d}{dx},x^3\right]
\Biggr]\Psi(x)
\label{Qmat3}
\end{eqnarray}
with $\Phi(x)=\tilde\Phi(x)/ \ell_P$, $\Psi(x)=\tilde\Psi(x)/ \ell_P$ and 
\begin{equation}
\bar q=2\beta(\bar a)\quad,\quad
\omega^2=-\frac{\left(\frac{d^2\beta(a)}{da^2}\right)_{a=\bar a}}
{\beta(\bar a)}>0\,,\label{defomega}
\end{equation}
\begin{equation}
c=\frac{\left(\frac{d^3\beta(a)}{da^3}\right)_{a=\bar a}}
{2\beta(\bar a)}\quad,\quad
d=\frac{\left(\frac{d^4\beta(a)}{da^4}\right)_{a=\bar a}}
{6\beta(\bar a)}\,.
\end{equation}
In calculating the r.h.s. of Eq.~(\ref{Qmat3}) we have repeatedly performed
integrations by parts and assumed the boundary terms to vanish. 
Introducing now the operators
\begin{equation}
\hat p=\frac{\ell_P^2}{i}\frac{d}{dx}\quad,\quad\hat x=x
\label{pxop}
\end{equation}
one easily reads off the effective operator expression
\begin{eqnarray}
& & \hat Q(\hat p,\hat x)
=\bar q\Biggl[1-\frac{\hat p^2}{2}-\frac{\omega^2}{2}\hat x^2
+\frac{c}{3}\hat x^3\nonumber\\
& & \qquad\qquad+\frac{d}{4}\hat x^4+\frac{\omega^2}{8}
\left(\hat x^2\hat p^2+\hat p^2\hat x^2\right)
+\frac{\hat p^4}{24}\nonumber\\
& & \qquad\qquad
+i\frac{\omega^2}{2}\left[\hat p,\hat x^2\right]
+i\frac{c}{3}\left[\hat p,\hat x^3\right]
\Biggr]\,.
\label{Qop}
\end{eqnarray}
This result extends the findings of Ref.~\cite{Schliemann13} to
higher corrections in the operators $\hat p$, $\hat x$.
The contribution in Eq.~(\ref{Qmat3}) involving only derivatives with
respect to $x$ can be viewed as the result of a continuum approximation
according to
\begin{eqnarray}
& & \langle a+\ell_P^2|\Psi\rangle+\langle a-\ell_P^2|\Psi\rangle-
2\langle a|\Psi\rangle\nonumber\\
& & \qquad\approx
\ell_P^4\frac{d^2\tilde\Psi(x)}{dx^2}
+\frac{\ell_P^8}{12}\frac{d^4\tilde\Psi(x)}{dx^4}\,.
\end {eqnarray}
Note also that the symmetric operator ordering in the last term of
the second line in Eq.~(\ref{Qmat3}) (i.e. the middle contribution
in the second line in Eq.~(\ref{Qop})) emerges from the calculation and
not an additional assumption.

As a result, the operator (\ref{Qop}) perfectly matches
the classical expression (\ref{qexpmax}) taking into account the correct
operator ordering and the vanishing of the commutators
\begin{equation}
\left[\hat p,\hat x^2\right]=-2i\ell_P^2\hat x\quad,\quad
\left[\hat p,\hat x^3\right]=-3i\ell_P^2\hat x^2
\end{equation}
which are indeed small compared to the other contributions in
(\ref{Qop}) as they are proportional to $\hbar$. 
Alternatively, the matrix elements of such commutators can be viewed to
be of higher order in derivatives since
\begin{eqnarray}
& & \frac{\omega^2}{2}\int dx\Phi^*(x)\left[\hat p,\hat x^2\right]\Psi(x)
\nonumber\\
& & \qquad=\frac{\omega^2}{2}
\int dx\left(\left(\hat p\Phi\right)^*\hat x^2\Psi
-\Phi^*\hat x^2\left(\hat p\Psi\right)\right)\,,
\end{eqnarray} 
where the r.h.s contains in total three derivatives with respect to $a$ or $x$.
Finally the coefficients in the expansions (\ref{Qop}) and (\ref{qexpmax})
obviously coincide in the limit of large quantum volumes, 
$\bar a\gg\ell_P^2$. In summary, up to the commutators discussed above, 
the operator (\ref{Qop}) is the result of the canonical quantization
of the classical expression (\ref{qexpmax}) via the standard
operator replacement (\ref{pxop}).

When concentrating on the quadratic contributions in Eq.~(\ref{Qop})
one recovers the harmonic-oscillator expression of Ref.~\cite{Schliemann13},
\begin{equation}
\hat Q_{\rm osc}(\hat p,\hat x)
=\bar q\left[1-\left(\frac{\hat p^2}{2}+\frac{\omega^2}{2}\hat x^2\right)
\right]\label{Qosc1}
\end{equation}
with eigenvalues
\begin{equation}
q_n^{\rm osc}=\bar q\left(1-\ell_P^2\omega(n+1/2)\right)\,.
\label{oscev}
\end{equation}
and corresponding eigenfunctions
\begin{equation}
\psi_n(x;\omega)
=\sqrt{\frac{1}{n!2^n}\sqrt{\frac{\omega}{\pi\ell_P^2}}}
H_n(\sqrt{\omega}x/ \ell_P)
e^{-\frac{\omega}{2\ell_P^2} x^2}
\label{oscwf}
\end{equation}
where $H_n(x)$ are the usual Hermite polynomials. We note that $\omega$ has
dimension of inverse area while $\ell_P^2\omega$ is dimensionless and can be
computed via Eqs.~(\ref{defomega}) using $\alpha(k)$ given in
Eq.~(\ref{alpha}) instead of $\beta(a)$,
\begin{equation}
\ell_P^4\omega^2=-\frac{\left(\frac{d^2\alpha(k)}{dk^2}\right)_{k=\bar k}}
{\alpha(\bar k)}\,.
\end{equation}

\begin{figure}
\includegraphics[width=\columnwidth]{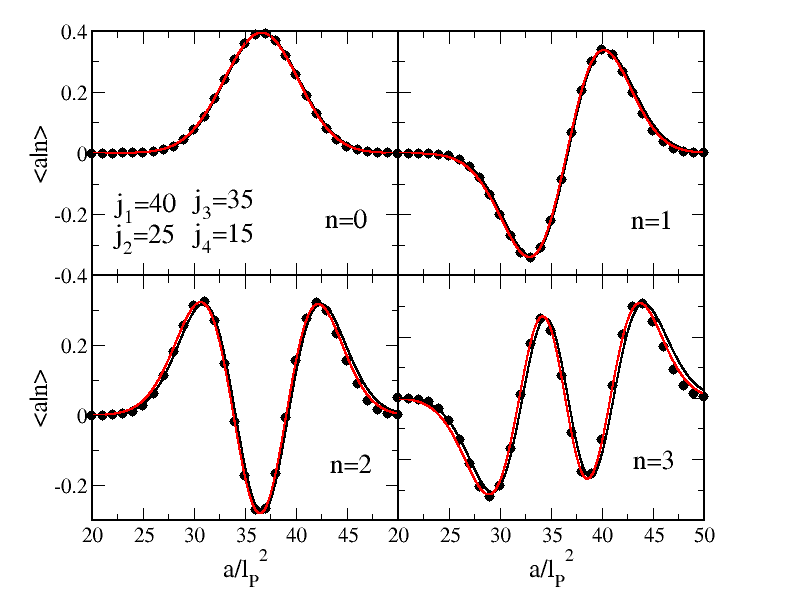}
\caption{The coefficients $\langle a|n\rangle$ (filled circles)
for small $n$ and a typical choice of angular momentum quantum numbers.
The black solid lines are the unperturbed oscillator wave functions 
$\psi_n^{(0)}(a-\bar a+\ell_P^2/2;\omega)$ (in units of $1/ \ell_P$)
given in Eq.~(\ref{oscwf}),
while the red lines show the eigenfunctions including the first-order
perturbation arising form the cubic term $c\hat x^3/3$ in Eq.~(\ref{Qop}).}
\label{fig1}
\end{figure}

\begin{figure}
\includegraphics[width=\columnwidth]{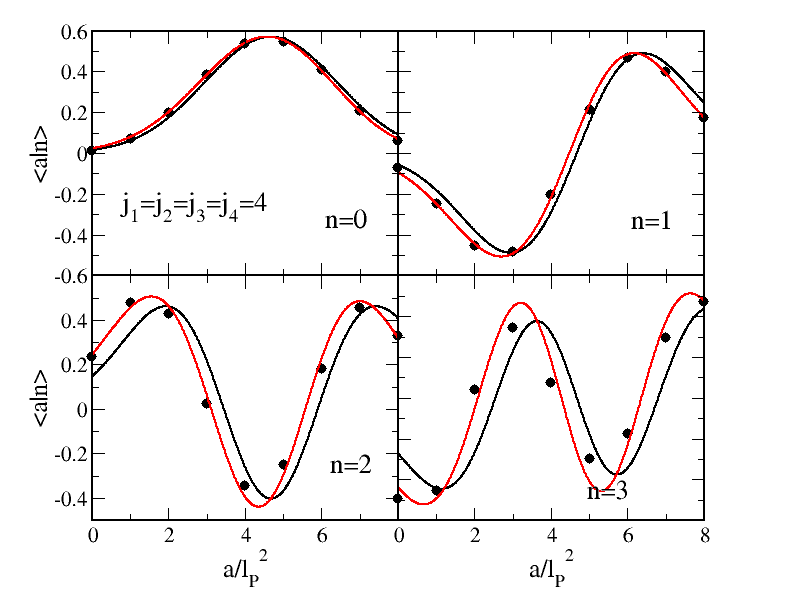}
\caption{The coefficients $\langle a|n\rangle$ (filled circles)
for small $n$ and $j_i\equiv 4$.
The black solid lines are the unperturbed oscillator wave functions 
$\psi_n^{(0)}(a-\bar a+\ell_P^2/2;\omega)$ (in units of $1/ \ell_P$)
given in Eq.~(\ref{oscwf}),
while the red lines show the eigenfunctions including the first-order
perturbation arising form the cubic term $c\hat x^3/3$ in Eq.~(\ref{Qop}).}
\label{fig2}
\end{figure}

As stated in Ref.~\cite{Schliemann13}, the expressions (\ref{oscev}) and
(\ref{oscwf}) are excellent approximations to the eigenstates and eigenvalues
of (square of the) the volume operator
for already intermediate lengths of the involved spins.
This fact is illustrated again in Fig.~\ref{fig1} for a typical typical 
choice of angular momentum quantum numbers all being of order
a few ten. In addition to Ref.~\cite{Schliemann13} we also plot there
the wave function within the lowest-order correction in Eq.~(\ref{Qop}
arising from $c\hat x^3/3$ accounted for by first-order perturbation
theory.
Fig.~\ref{fig2} shows similar data but for smaller spin lengths
$j_i\equiv 4$. Here the oscillator-like features of the wave functions
noticeably disappear with increasing $n$, and the corrections from cubic
term are clearly more substantial.
In both Figs.~\ref{fig1} and \ref{fig2} we have used the expression
$a-\bar a+\ell_P^2/2$ as the argument for the wave functions where the 
additional increment $\ell_P^2$ takes into account
that $\beta(\bar a)$ couples states of the form $|\bar a-\ell_P^2\rangle$
and $|\bar a\rangle$ and facilitates comparison with finite size data.
With increasing angular momentum quantum numbers, this shift becomes more
and more obsolete.

\subsubsection{Negative Eigenvalues of $\hat Q$}
\label{quanttetraneg}

So far we have concentrated on the large and positive eigenvalues of
the operator $\hat Q$. The regime of negative eigenvalues of large modulus
can be explored by canonically 
quantizing the classical expression (\ref{oscmin})
according to the standard recipe (\ref{pxop}),
\begin{equation}
\hat Q_{\rm osc}^{\prime}(\hat p,\hat x)
=-\bar q\left[1-\left(\frac{(\hat p-\pi)^2}{2}+\frac{\omega^2}{2}\hat x^2\right)
\right]\,,
\label{Qosc2}
\end{equation}
where we have put for simplicity $\tilde q=\bar q$, $\tilde\omega=\bar \omega$.
This operator is related to $Q_{\rm osc}$ given in Eq.~(\ref{Qosc1}) by
a gauge transformation along with a change in sign,
\begin{equation}
Q_{\rm osc}^{\prime}=-e^{i\pi x/ \ell_P^2}Q_{\rm osc}e^{-i\pi x/ \ell_P^2}
\label{Qosc3}
\end{equation}
such that the eigenfunctions are related by
\begin{equation}
\psi_n^{\prime}(x)=e^{i\pi x/ \ell_P^2}\psi_n(x)\,,
\label{oscwf2}
\end{equation}
where the phase factor mimics the change in sign stated in the 
strict relation
(\ref{signchange}) between eigenvectors of $\hat Q$ to eigenvalues
differing in sign only. In fact, based on this analogy, Eq.~(\ref{oscwf2})
and, as a consequence, Eqs.~(\ref{Qosc2}),(\ref{Qosc3}) have already been given
in Ref.~\cite{Schliemann13}. Here we have provided a more profound derivation
based on the canonical quantization of the classical expression (\ref{oscmin}).

\subsubsection{Canonical Operators in the Discrete Case}

In the operators (\ref{pxop}) the variable $x$ (and, in turn, $a$) is
considered to be a continuous quantity. Therefore the question remains how
to possibly construct a pair of canonical operators retaining the 
discrete character of $a=k/ \ell_P^2$ with $k$ being (half-)integer.
As a step towards this goal we propose the operators
\begin{eqnarray}
\hat A & = & \sum_{a=a^{\rm min}}^{a^{\rm max}}a|a\rangle\langle a|\,.
\label{Aop}\\
\hat P & = & \frac{i}{2}\sum_{a=a^{\rm min}+\ell_P^2}^{a^{\rm max}}
\left(|a\rangle\langle a-\ell_P^2|-|a-\ell_P^2\rangle\langle a|\right)
\label{Pop}
\end{eqnarray}
fulfilling
\begin{eqnarray}
& & \left[\hat P,\hat A\right]\nonumber\\
& & \quad=\frac{\ell_P^2}{2i}
\sum_{a=a^{\rm min}+\ell_P^2}^{a^{\rm max}}
\left(|a\rangle\langle a-\ell_P^2|+|a-\ell_P^2\rangle\langle a|\right)\,.
\label{PAcomm}
\end{eqnarray}
For large volumes, the r.h.s. approaches the unit operator acting on states
whose components vary only little on the scale set by $\ell_P^2$,
\begin{eqnarray}
\left\langle\Phi\left|\left[\hat P,\hat A\right]\right|\Psi\right\rangle
& \approx & 
\sum_{a=a^{\rm min}}^{a^{\rm max}}\langle\Phi|a\rangle\langle a|\Psi\rangle
\label{PAcommmat}\\
& = & \frac{\ell_P^2}{i}\langle\Phi|\Psi\rangle\,.
\end{eqnarray}
In fact, the expression (\ref{Pop}) is obviously a discretization
of a differential operator. However, as such discretizations are by no means
unique, the question remains open whether there are operators
$\hat P^{\prime}$, $\hat A^{\prime}$ 
which (i) act on the original discretely-labeled
quantum states, (ii) turn into $\hat p$, $\hat x$ at large volumes,
and (iii) fulfill
\begin{equation}
\left[\hat P^{\prime},\hat A^{\prime}\right]=\frac{\ell_P^2}{i}
\end{equation}
as an exact equation on the entire Hilbert space.

\subsection{General Polyhedra}
\label{quantpolygen}

In full analogy to the Kapovich-Millson variables we define for a quantum
polyhedron of $N$ faces (angular momenta) the operators
\begin{equation}
\hat{\vec k}_i=\sum_{j=1}^{i+1}\hat{\vec j}_j
\end{equation}
for $i\in\{1,\dots,N-3\}$. As the squares of these quantities commute with
each other,
\begin{equation}
\left[\hat{\vec k}_i^2,\hat{\vec k}_j^2\right]=0\,,
\end{equation}
orthonormal basis states of the Hilbert space can be labeled by
quantum numbers $k_i$ according to
\begin{equation}
\hat{\vec k}_i^2|k_1\dots k_{N-3}\rangle=k_i(k_i+1)|k_1\dots k_{N-3}\rangle\,.
\end{equation}
The closure relation (\ref{clorel}) translates to
\begin{equation}
\sum_{i=1}^N\hat{\vec j}_i|k_1\dots k_{N-3}\rangle=0\,,
\label{singcond}
\end{equation}
i.e. the angular momentum operator $\hat{\vec k}_{N-3}$ couples with
the remaining spins $\hat{\vec j}_{N-1}$, $\hat{\vec j}_N$ 
to a total singlet implying
$k_{N-3}^{\rm min}\leq k_{N-3}\leq k_{N-3}^{\rm max}$ with
\begin{eqnarray}
k_{N-3}^{\rm min} & \geq & |j_{N-1}-j_N|\,,
\label{kcond1}\\
k_{N-3}^{\rm max} & \leq & j_{N-1}+j_N\,,
\label{kcond2}
\end{eqnarray}
Consider now two total singlet states with $k_i=k_i^{(1)}$ and
$k_i=k_i^{(2)}$, $i<N-3$, $k_i^{(1)}<k_i^{(2)}$ and all other quantum
numbers $k_j$, $j\neq i$ identical. Then states
with $k_i=k_i^{(1)}+1,\dots,k_i^{(2)}-1$ (and all other $k_j$ the same as before)
are also singlets, since $\hat{\vec k}_{i-1}$ and $\hat{\vec j}_{i+1}$ 
can couple to these
values of $k_i$, and $\hat{\vec k}_i$ with the above quantum numbers
and $\hat{\vec j}_{i+2}$ can couple to the given
value of $k_{i+1}$. Thus, also the other quantum
quantum numbers $k_i$, $i<N-3$, vary within intervals,
$k_i^{\rm min}\leq k_i\leq k_i^{\rm max}$, and the representation theory
of the angular momentum algebra implies
\begin{eqnarray}
k_i^{\rm min} & \geq & \max\{k_{i-1}^{\rm min}-j_{i+1},0,
j_{i+1}-k_{i-1}^{\rm max}\}\,,
\label{kcond3}\\
k_i^{\rm max} & \leq & k_{i-1}+j_{i+1}
\label{kcond4}
\end{eqnarray}
with $k_0=j_1$ for $i=1$. 
Without the additional
conditions (\ref{kcond1}),(\ref{kcond2}) the inequalities
(\ref{kcond3}),(\ref{kcond4}) would hold as equalities.
We note, however, that the structure of
the quantum numbers $k_i$ is in general quite complex. For instance, the
limiting values $k_i^{\rm min}$, $k_i^{\rm max}$ can depend on other
quantum numbers $k_j$, $j\neq i$, and the entire structure
depends obviously also on the coupling scheme, i.e. the labeling
of the operators $j_1,\dots,j_N$. A very simple example is
provided by the case $N=4$ in Eqs.~(\ref{tetkcond1}),(\ref{tetkcond2});
for $N>4$ explicit expressions for $k_i^{\rm min}$, $k_i^{\rm max}$ become
increasingly tedious.

Now rescaling the quantum numbers $k_i$ to dimensionful quantities as
in section \ref{quanttetrarescale}, $k_i\mapsto a_i=k_i\ell_P^2$,
we define analogous to Eqs.~(\ref{Aop}),(\ref{Pop}) the operators
\begin{eqnarray}
\hat A_i & = & \sum_{a_1\dots a_{N-3}}a_i|a_1\dots a_{N-3}\rangle
\langle a_1\dots a_{N-3}|\,,\\
\hat P_i & = & \frac{i}{2}\sum_{a_1\dots a_{N-3}}\Bigl(
|a_1\dots a_{N-3}\rangle\langle a_1\dots(a_i-\ell_P^2)\dots a_{N-3}|\nonumber\\
& & -|a_1\dots(a_i-\ell_P^2)\dots a_{N-3}\rangle\langle a_1\dots a_{N-3}|\Bigr)
\end{eqnarray}
fulfilling the commutation relations
\begin{eqnarray}
& & \left[\hat P_i,\hat A_j\right]=\frac{\delta_{ij}\ell_P^2}{2i}
\nonumber\\
& & \quad\cdot\sum_{a_1\dots a_{N-3}}\Bigl(
|a_1\dots a_{N-3}\rangle\langle a_1\dots(a_i-\ell_P^2)\dots a_{N-3}|\nonumber\\
& & \quad+|a_1\dots(a_i-\ell_P^2)\dots a_{N-3}\rangle\langle a_1\dots a_{N-3}
|\Bigr)\,.
\end{eqnarray}
In the limit of large volumes, the above r.h.s. approaches the unit operator 
in just the same way as in Eq.~(\ref{PAcomm}).

\section{Conclusions and Outlook}

The investigation of the semiclassical limit of Loop Quantum Gravity is one
of the key issues in that approach towards a quantum theory of gravitation.
In the present work we have focussed semiclassical properties of quantum
polyhedra. Regarding tetrahedra,
as their simplest examples, Eqs.~(\ref{defvwop})-(\ref{projec})
provide operator analogs of classical geometric relations
for tetrahedra. These classical relations are key ingredient to the
Bohr-Sommerfeld analysis of Refs.~\cite{Bianchi11b,Bianchi12}.

The expansion of the classical volume squared in up to fourth order
in the canonical variables are given in Eqs.~(\ref{qexpmax}, (\ref{qexpmin}).
we have explicitly established the connection between
a canonical quantization of these expressions with a recent
wave function based
approach by the present author \cite{Schliemann13}
to the large-volume sector of the quantum system.
In the leading order both routes concur 
yielding a quantum harmonic oscillator
as an effective description for the (square of the) volume operator.
As regards higher orders, the approach of Ref.~\cite{Schliemann13}
leads to additional corrections in terms of commutators which are naturally
absent in the classical expressions.
Including the third-order correction perturbatively leads to improvements
of the approximate wave functions. In fact, it is a distinctive feature of the
present work (and Ref.~\cite{Schliemann13}) that it addresses not only the
{\em eigenvalues} of the volume operator squared, but also provides 
very accurate approximations to the {\em eigenstates}.
Furthermore, the comparison of both methods
leads also to a full wave function description of the eigenstates 
of negative eigenvalues of large modulus, a result which could only be
conjectured in Ref.~\cite{Schliemann13}. 

Differently from previous formulations,
the position variable used here is chosen to have dimension of (Planck) length
squared, $\ell_P^2=\hbar G/c^3$. 
This definitional detail is by no means necessary
but facilitates the identification of quantum corrections. The ultimate
reason for the latter observation is the fact that Planck\rq s constant
$\hbar$ itself is dimensionful.

A further interesting point is the zero eigenvalue occurring for tetrahedra
with odd Hilbert space dimension $d$ where the eigenstate can be given, up to
normalization, in a closed form \cite{Brunnemann06}. Moreover, the 
Bohr-Sommerfeld quantization carried out by Bianchi and Haggard 
\cite{Bianchi11b,Bianchi12} yields also surprisingly accurate results
for eigenvalues of such small modulus. Thus the question arises whether
the eigenstates corresponding to zero eigenvalues can also be cast, for large
angular momenta $j_i\gg 1$, in a wave function of a continuous variable.

An important step towards extending the results for the tetrahedron to
higher polyhedra is to express their volume in terms of canonical variables.
In appendix \ref{penta} we have achieved this goal for the case of pentahedra.
However, the resulting expressions are particularly lengthy and complex
such that further practical progress seems to require dedicated numerics
and/or extensive but judicious use of computer algebra, 
which is beyond the scope of the present investigation.

For general quantum polyhedra described by discrete angular momentum
quantum numbers we have formulated a set of
quantum operators fulfilling in the semiclassical regime 
the standard commutation relations
between momentum and position. 
Indeed a major challenge is of course the analysis of the 
volume operator(s) for
higher polyhedra. Results towards this goal were obtained in 
Refs.~\cite{Haggard13,Coleman-Smith13} for pentahedra, and for the
general case the operators constructed in the present paper in analogy to the
Kapovich-Millson variables of the classical phase space might, although fairly
straightforward, 
provide a useful step. Yet another possible route for generalizing the
present investigations is to study polytopes in higher dimensions which also
allow for a description in terms of SU(2) intertwiners \cite{Bodendorfer13}.

\appendix 

\section{The Classical Pentahedron and Canonical Variabels}
\label{penta}

In this appendix we discuss the volume of a classical pentahedron in terms of
Kapovich-Millson variables. We concentrate on the dominant type of
pentahedra which define the submanifold of maximal dimension in the
phase space of the variables $\vec A_i$ \cite{Bianchi11a}.  
These pentahedra are trigonal prisms whereas a subdomiant type is given
by pyramidal pentahedra. These can be generated form the former type by
collapsing an edge connecting the two trigonal faces onto a single point
and form therefore a submanifold of lower dimension.

\subsection{Volume}

\begin{figure}
\includegraphics[width=6cm]{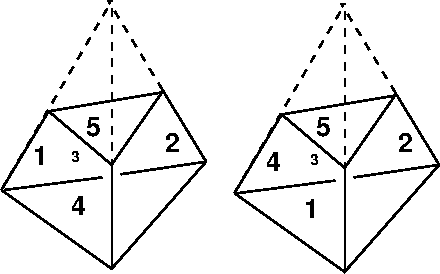}
\caption{A pentahedron of dominant type (trigonal prism) extended to a
tetrahedron. Two different labelings of faces are shown.}
\label{fig3}
\end{figure}
An expression for the volume of a trigonal prism has been devised by
Haggard \cite{Haggard13} starting from the observation that such an
object can always be extended to a tetrahedron. Using the labeling on the
left of Fig.~\ref{fig3} this extended body fulfills the closure relation
\begin{equation}
\alpha\vec A_1+\beta\vec A_2+\gamma\vec A_3+\vec A_4=0\,,
\label{scaling1}
\end{equation}
and by projecting this equation onto appropriate cross products the above
coefficients are easily obtained as
\begin{equation}
\alpha=-\frac{W_{234}}{W_{123}}\,,\quad\beta=\frac{W_{134}}{W_{123}}\,,\quad
\gamma=-\frac{W_{124}}{W_{123}}
\label{coeff1}
\end{equation}
with $W_{ijk}=\vec A_i\cdot(\vec A_j\times\vec A_k)$.
The volume can now be expressed as the difference ot two tetrahedral volumes,
\begin{equation}
V=\frac{\sqrt{2}}{3}\left(
\sqrt{\alpha\beta\gamma}-\sqrt{(\alpha-1)(\beta-1)(\gamma-1)}\right)
\sqrt{W_{123}}\,.
\label{pentvol}
\end{equation}

\subsection{Canonical Variables}

Our goal is now to express the scaling coefficients (\ref{coeff1})
occuring along $W_{123}$ in Eq.~(\ref{pentvol}) in terms of standard
Kapovich-Millson variables. According to the presrciption given in
section \ref{clpolyKP} we define
\begin{eqnarray}
\vec p_1 & = & \vec A_1+\vec A_2\,,
\label{defp1}\\
\vec p_2 & = & \vec A_1+\vec A_2+\vec A_3
\label{defp2}
\end{eqnarray}
and 
\begin{eqnarray}
\vec v_1 & = & \vec p_1\times\vec A_2\,,
\label{defv1}\\
\vec w_1 & = & \vec v_2=\vec p_2\times \vec A_3\,,
\label{defw1}\\
\vec w_2 & = & \vec p_2\times \vec A_4
\label{defw2}\\
\end{eqnarray}
such that the variables $q_i$, $i=1,2$, conjugate to $p_i:=|\vec p_i|$
are the angles between $\vec v_i$, $\vec w_i$. Using the definition 
(\ref{defDel}) one has
\begin{eqnarray}
|\vec v_1| & = & 2\Delta(p_1,A_1,A_2)\,,
\label{modv1}\\
|\vec w_1| & = & |\vec v_2|=2\Delta(p_1,p_2,A_3)\,,
\label{modw1}\\
|\vec w_2| & = & 2\Delta(p_2,A_4,A_5)\,.
\label{modw2}
\end{eqnarray}
Moreover, the relations
\begin{eqnarray}
\vec v_1\times \vec w_1 & = & \vec p_1W_{123}\,,
\label{v1xw1}\\
\vec v_2\times \vec w_2 & = & \vec p_1\left(W_{134}+W_{234}\right)
\label{v2xw2}
\end{eqnarray}
allow us to achieve a part of our task in a comparatively compact
manner,
\begin{eqnarray}
W_{123} & = & \frac{4\Delta(p_1,A_1,A_2)\Delta(p_1,p_2,A_3)\sin q_1}{p_1}\,,
\label{W123}\\
\beta-\alpha & = & \frac{\Delta(p_2,A_4,A_5)p_1\sin q_2}
{\Delta(p_1,A_1,A_2)p_2\sin q_1}\,.
\label{b-a}
\end{eqnarray}
Unfortunately, the remaining quantities entering the volume (\ref{pentvol})
will turn out to lead to clearly lengthier expressions. To compute them, we
shall not aim at accessing further triple products $W_{ijk}$ directly but
rather project the closure relation (\ref{scaling1}) onto $\vec p_i$,
\begin{eqnarray}
\alpha\vec p_1\cdot\vec A_1+\beta\vec p_1\cdot\vec A_2
+\gamma\vec p_1\cdot\vec A_3 & = & -\vec p_1\cdot\vec A_4\,,
\label{prop1}\\
\alpha\vec p_2\cdot\vec A_1+\beta\vec p_2\cdot\vec A_2
+\gamma\vec p_2\cdot\vec A_3 & = & -\vec p_2\cdot\vec A_4\,,
\label{prop2}
\end{eqnarray}
providing two further equations for $\alpha$, $\beta$, $\gamma$
with coeffcients we will determine now.

From Eqs.~(\ref{defp1}),(\ref{defp2}) one easily finds
\begin{eqnarray}
\vec p_1\cdot\vec A_{1/2} & = & \frac{1}{2}
\left(p_1^2\pm\left(A_1^2-A_2^2\right)\right)\,,
\label{p1A1}\\
\vec p_{1/2}\cdot\vec A_3 & = & \frac{1}{2}\left(p_2^2-p_1^2\mp A_3^2\right)\,,
\label{p1A3}
\end{eqnarray}
along with
\begin{equation}
\vec p_1\cdot\vec p_2=\frac{1}{2}\left(p_1^2+p_1^2-A_3^2\right)\,,
\label{p1p2}
\end{equation} 
and, via the closure relation (\ref{clorel}),
\begin{equation}
\vec p_2\cdot\vec A_4=-\frac{1}{2}\left(p_2^2+A_4^2-A_5^2\right)\,.
\label{p2A4}
\end{equation}

In order to determine $\vec p_2\cdot\vec A_{1/2}$ we calculate,
using Eq.~(\ref{p1A1}),
\begin{eqnarray}
\vec v_1\cdot\vec w_1 & = & \left(\left(\vec A_1\times\vec A_2\right)
\times\vec p_1\right)\cdot\vec A_3\nonumber\\
& = & -\vec A_1\cdot\vec A_3\left(\vec p_1\cdot\vec A_2\right)
+\vec A_2\cdot\vec A_3\left(\vec p_1\cdot\vec A_1\right)\nonumber\\
& = &  \frac{1}{2}\left(A_1^2-A_2^2\right)\vec p_1\cdot\vec A_3\nonumber\\
& & \qquad-\frac{1}{2}p_1^2\left(\vec A_1-\vec A_2\right)\cdot\vec A_3
\end{eqnarray}
such that, taking into account q.~(\ref{p1A3})
\begin{eqnarray}
\vec A_{1/2}\cdot\vec A_3 & = &
\frac{1}{4}\left(1\pm\frac{A_1^2-A_2^2}{p_1^2}\right)
\left(p_1^2\pm\left(A_1^2-A_2^2\right)\right)\nonumber\\
& &  \qquad\mp\frac{\vec v_1\cdot\vec w_1}{p_1^2}
\end{eqnarray}
and finally
\begin{eqnarray}
\vec p_2\cdot\vec A_{1/2} & = & \vec p_1\cdot\vec A_{1/2}
+\vec A_3\cdot\vec A_{1/2}\nonumber\\
& = & \frac{1}{4}\left(p_1^2+p_2^2+A_3^2\right)\nonumber\\
& & \quad\pm\left(A_1^2-A_2^2\right)
\frac{p_1^2+p_2^2+A_3^2}{4p_1^2}\nonumber\\
& & \quad\mp\frac{\vec v_1\cdot\vec w_1}{p_1^2}
\label{p2A1}
\end{eqnarray}
where
\begin{equation}
\vec v_1\cdot\vec w_1=4\Delta(p_1,A_1,A_2)\Delta(p_1,p_2,A_3)\cos q_1\,.
\end{equation}
Similarly, one finds
\begin{eqnarray}
\vec v_2\cdot\vec w_2 & = & -p_2^2\vec p_1\cdot\vec A_4\nonumber\\
& & \qquad+\frac{1}{2}\left(p_2^2+p_1^2-A_3^2\right)\vec p_2\cdot\vec A_4\,,
\end{eqnarray}
which yields in combination with Eq.~(\ref{p2A4})
\begin{eqnarray}
\vec p_1\cdot\vec A_4 & = & \frac{p_1^2+p_2^2-A_3^2}{4p_2^2}
\left(p_2^2+A_4^2-A_5^2\right)\nonumber\\
& & \quad-\frac{\vec v_2\cdot\vec w_2}{p_2^2}
\label{p1A4}
\end{eqnarray}
with
\begin{equation}
\vec v_2\cdot\vec w_2=4\Delta(p_1,p_2,A_3)\Delta(p_2,A_4,A_5)\cos q_2\,.
\end{equation}
Thus we have expressed all scalar products occurring in 
Eqs.~(\ref{prop1}),(\ref{prop2}) in terms of the canonical variables
$p_i$, $q_i$. Taking into account Eq.~(\ref{b-a}), these relations can now be
formulated as
\begin{equation}
M(p_1,p_2)\left(
\begin{array}{c}
\alpha+\beta \\
\gamma
\end{array}
\right)
=\left(
\begin{array}{c}
F(p_i,q_i) \\
G(p_i,q_i)
\end{array}
\right)
\label{a+bg}
\end{equation}
with
\begin{equation}
M(p_1,p_2)=\left(
\begin{array}{cc}
p_1^2 & p_2^2-p_1^2-A_3^2 \\
\frac{1}{2}\left(p_2^2+p_1^2+A_3^2\right) & p_2^2-p_1^2+A_3^2
\label{defM}
\end{array}
\right)
\end{equation}
and
\begin{eqnarray}
F(p_i,q_i) & = & \frac{\Delta(p_2,A_4,A_5)p_1\sin q_2}
{\Delta(p_1,A_1,A_2)p_2\sin q_1}\left(A_1^2-A_2^2\right)\nonumber\\
& & \quad-\frac{p_1^2+p_2^2-A_3^2}{2p_2^2}
\left(p_2^2+A_4^2-A_5^2\right)\nonumber\\
& & \quad+2\frac{\vec v_2\cdot\vec w_2}{p_2^2}\,,
\label{defF}\\
G(p_i,q_i) & = & \frac{\Delta(p_2,A_4,A_5)p_1\sin q_2}
{\Delta(p_1,A_1,A_2)p_2\sin q_1}\Biggl[
-2\frac{\vec v_1\cdot\vec w_1}{p_1^2}\nonumber\\
& & \qquad\qquad+\left(A_1^2-A_2^2\right)
\frac{p_1^2+p_2^2+A_3^2}{2p_1^2}\Biggr]\nonumber\\
& & \quad+p_2^2+A_4^2-A_5^2\,.
\end{eqnarray}
Now, inverting the $2\times 2$-matrix (\ref{defM}) and using again
Eq.~(\ref{b-a}) one can explicitly solve for the scaling coefficients
$\alpha$, $\beta$, $\gamma$. This procedure, however, will obviously result in
forbiddingly lengthy and complicated
expressions, which is mainly due to the cumbersome structures
in on the r.h.s of Eq.~(\ref{a+bg}). We note that these expressions 
considerably simplify for $A_1=A_2$ and $A_4=A_5$, i.e. if the two traingular
faces and two of the other faces have pairwise the same area. Then one has
\begin{eqnarray}
F(p_i,q_i) & = & -\frac{1}{2}\left(p_1^2+p_2^2-A_3^2\right)
+2\frac{\vec v_2\cdot\vec w_2}{p_2^2}\,,\\
G(p_i,q_i) & = & -\frac{8\Delta(p_1,p_2,A_3)\Delta(p_2,A_4,A_4)}
{p_1p_2}\frac{\sin q_2}{\tan q_1}\nonumber\\
& & \qquad+p_2^2\,.
\end{eqnarray}
However, also these quantities seem to be too complicated to allow
for further practical analytical progress towards, e.g., the extrema of 
the pentahedral volume and expansions around them.

\subsection{Relabelings and Alternative Variables}

One might suspect that simpler expressions for the pentahedral volume
can be obtained by a judicious alternative choice of the canonical variables.
For example, an apparently more symmetric arrangement would be to
couple the trigonal faces separately with other faces to canonical momenta.
Using the labeling given on the right of Fig.~\ref{fig3}, this means to
use the definitions (\ref{defp1} ),(\ref{defv1}),(\ref{defw1}) as before
and put
\begin{eqnarray}
\vec p^{\prime}_2 & = & \vec A_4+\vec A_5\,,\\
\vec v^{\prime}_2 & = & \vec p^{\prime}_2\times\vec A_4\,,\\
\vec w^{\prime}_2 & = & \vec p^{\prime}_2\times\vec A_3\,.\\
\end{eqnarray}
However, the closure relation (\ref{clorel}) immediately tells that
\begin{equation}
\vec p^{\prime}_2 =-\vec p_2\,,\quad
\vec v^{\prime}_2 =-\vec w_2\,,\quad
\vec w^{\prime}_2 =-\vec v_2\,.
\end{equation}
Thus, up to inessential signs, we end up with the same canonical variables as
before. Moreover, the closure realtion for the extended tetrahedral volume
reads now
\begin{equation}
\vec A_1+\alpha^{\prime}\vec A_2+\beta^{\prime}\vec A_3+\gamma^{\prime}\vec A_4=0\,,
\label{scaling2}
\end{equation}
where the new scaling coefficients can be expressed in terms of the old ones
(\ref{coeff1}) as \cite{Haggard13}
\begin{equation}
\alpha^{\prime}=\frac{\beta}{\alpha}\,,\quad
\beta^{\prime}=\frac{\gamma}{\alpha}\,,\quad
\gamma^{\prime}=\frac{1}{\alpha}\,.
\label{coeff2}
\end{equation}
As a result, we encounter very similar technical difficulties. Furthermore,
in Ref.~\cite{Haggard13} an exhaustive list of pentahedral face labelings
and corresponding scaling coefficients has been given. Inspecting these
results does also not give rise to the hope that such a change of variables
will lead to substantially simpler expressions for the volume.

\end{document}